\documentstyle{article}
\begin{document}
%\baselineskip=0.5 in
%\def\baselinestretch{2}
%\baselineskip=24.0pt
%\large
{}\hfill {\Large \bf IFT/05/98}
\vskip.5in
\centerline{\Large {\bf The Living Network of Schools}}
\vskip.1in
\centerline{\Large {\bf Owned by Teachers and Students}}
\vskip.1in
\centerline{April 1, 1998}

\vskip .3in
\centerline{Stanis{\l}aw D. G{\l}azek}
\vskip .1in
\centerline{Institute of Theoretical Physics, Warsaw University}
\centerline{ul. Ho{\.z}a 69, 00-681 Warsaw}

\vskip.3in
\centerline{\bf Abstract}
\vskip.1in

This paper describes a network of teachers and students who form a living system
of education at all levels.  Organization of schools is based on new principles.
One person can be a teacher in one area or activity and a student in another.
Schools are owned and governed by the teachers and students.  The system is
powered at all levels equally by the will of students to learn and the will of
teachers to learn and to share their expertise with students. 

We describe main processes and structural principles of the network.  The key
processes are the process of learning by inquiry and the processes of design and
learning by redesign.  We also describe steps required to initiate the network
growth process from small scale seeds.  This avoids wasting human resources and 
money on a large scale. The first step we suggest for the teams of teachers and 
researchers who are interested in building the network is studying a bit of 
basic physics by inquiry using specially designed and well tested materials.  

The network is economically sound.  We distinguish the economy of the network
because we claim that the freedom and safety of learning and teaching processes
can be based only on the financial independence of teachers who gained their
independence as a result of developing and using these processes.  The system is
designed to ensure highest quality in all respects.  The design described here
is provided as illustration for the underlying principles and their implications
rather than as the ultimate structure.  In fact, the living network is expected
to evolve and adapt efficiently.

%\newpage
\vskip.5in
%\centerline{\bf CONTENTS:}
{\bf CONTENTS:}
%\vskip.3in
\vskip.1in

\parbox{15cm}{
{1. Introduction \dotfill {\it The lion claw and the educational trade mark}  ~~~  2 }
\vskip.05in
%\vskip.2in
{2. Structure of the network \dotfill {\it The super-conducting wiring of learning by inquiry} ~~~ 4 }
\vskip.05in
%\vskip.2in
{3. Ownership principles \dotfill {\it The rocket riders build and share the spaceship}  
~~~ 7 }
\vskip.05in
%\vskip.2in
{4. Governance principles \dotfill {\it Brain cells know their responsibilities} ~~~ 8 }
\vskip.05in
%\vskip.2in
{5. Sources of income and finance management \dotfill {\it Buy your pass to the fastest 
lane} ~~~ 9 }
\vskip.05in
%\vskip.2in
{6. Assessment and recognition \dotfill {\it The credit guaranteed in gold} ~~~ 10}
\vskip.05in
%\vskip.2in
{7. Archives \dotfill {\it You may know what is going on} ~~~ 11}
\vskip.05in
%\vskip.2in
{8. Preparation, design, launch, feedback and redesign \dotfill {\it Neverending adventure} ~~~ 11}
\vskip.05in
%\vskip.2in
{9. Community and state support \dotfill {\it They line up to help} ~~~  10}
\vskip.05in
%\vskip.2in
{10. Seeds and timing \dotfill {\it A sequoia seed is billions times smaller than the tree} 
~~~ 11}
\vskip.05in
%\vskip.2in
{11. Initial business proposal \dotfill {\it Are you ready?} ~~~ 12}
\vskip.05in
%\vskip.2in
{12. Communication \dotfill {\it We enable people to learn effectively} ~~~ 12}
\vskip.05in
%\vskip.2in
{13. Conclusion \dotfill {\it Will universities lead in future?} ~~~ 13}}

\newpage

{\bf 1. INTRODUCTION}
\vskip.1in

It is bad that our educational system does not teach students how to learn
effectively.  It is bad students are often lost and not interested in learning
at school.  But it is very bad that teachers at all levels get used to thinking
they cannot change much in the way they teach.  First of all, we teach the way
we were taught.  Secondly, it is very hard to make changes.  We want to reach
our students' minds but we feel we have to do it in the environment which we
have little influence on as individuals.  Most importantly, the environment is
not safe for experimentation on better ways of teaching. We know we have to complete 
an overwhelmingly difficult program.  We know we have unacceptably short time to 
do so.  We have little freedom to make choices.  And we will be criticized if we 
admit our students do not learn much.  In fact, we do not even know how bad and how 
little they learn, although we certainly know it is not enough.  More interestingly, 
we hesitate to work on measuring how much they learn. We do not have time to do that. 
And it would be hard to accept to everybody in the system that our students do 
not learn how to think critically, how to learn effectively and how to approach 
new problems requiring solution. But if we admitted we do not know how to teach 
effectively we would question our competence and jeopardize our living.

Assuming that the overarching goal of education is to train people in learning
effectively so that they are able to learn and change throughout their whole
life, one is faced with the question:  What needs to be changed in our current
concept of teaching?  For example, it is clear that a child learns very
efficiently when it feels safe and when it can experiment.  It can touch, taste,
break and throw things, make its mother angry, etc.  The child learns from
failure and pain or success and pleasure.  When such learning is wisely
instructed I call it learning by inquiry.  In mature form learning by inquiry is
the only way humans truly learn new things and become owners and users of what
they learned.  I claim teachers could learn how to teach effectively if they
worked in the environment safe for experimentation and had good examples to
study and patterns to discover.  Also, children will eagerly learn if
guided by a teacher who feels like a lion letting cubs play with food before
showing them how to hunt.  Teachers will be self-motivated to learn and
to teach by inquiry if the best sources to learn from are made available to them.
One such already existing source is physics, the most advanced science driven by
inquiry.  I will return to this point later.

If the freedom to learn is the prerequisite to teachers' action, on the one
hand, and the fear of rejection when attempting new approaches originates in the
fear of losing financial security in the current system, on the other hand,
then, the way out is to create a system in which successful attempts of teaching
in a better way will be rewarded with more financial independence for teachers.
This independence is the bottom line security condition which we do not have
satisfied and which we have to have satisfied to feel free to explore.  But our
safety has to result directly from our work, not from top-down orders.  The
reason is that the arbitrary top-down ruling may change at any time for some
reason and the temporarily existing safety will be gone, as if the lion went
away.  However, if teachers and students work in a self-conscious
network, which is in control of its income and spending and which is driven by
merit of the overarching goal of education and, if teachers generate their
income through their own action then, no arbitrary budget decisions could take
away from us our freedom to learn and teach by inquiry the best way we can.

In the network I envision, whenever we will see the excitement about learning
new things in the eyes of our students we will know we are making irreversible
progress.  We will be learning ourselves how to better the way we teach.  We
will engage in the work on creating new opportunities for achieving better
results more often.  We will be building our own system, and our own future.
Thus, the ownership principle becomes the basis for the overarching goal of
education.  We the teachers and students working in our own network become the
lions protecting our own development and future and we become the guarantors of
our freedom to learn by inquiry both the subject matters and how to teach better.

Being a teacher and observing teachers in schools of all levels one can notice
that the money a teacher earns for doing the job is not related to the depth of
learning experience provided to students.  On average, teachers are not in a
position to be entrepreneurial \cite{Drucker} in their schools and they do not
pursue quests like motivated professionals who create prosperity of their
disciplines \cite{Hughes}.  Teachers themselves do not achieve clear results
that could raise awe in students and motivate them to learning.  A tired teacher
at the end of the day is supposed to check a pile of poorly done homework and
accept the lack of future.  Little time is left to the teacher for personal
growth, gaining respect and winning intellectual freedom.  \cite{Covey} The lion
cub eats well after a good hunt.  Teachers eat the same no matter how well they
teach.  But the coupling I postulate between the results of teaching and
teacher's income must not be confused with using greed for money to serve
educational purposes.  That would certainly not work.

The money reward to teachers in contemporary society is essential for many
reasons.  The key one is that low-paid teachers cannot teach students about
achievements of the society.  For they know these achievements only second hand
and only as much as can be bought using very limited and already allocated
funds.  And currently even a modest attempt on the part of public school
teachers to change this situation may take away their basic income because the
system is not open to experimentation.  Moreover, if the high quality teacher's
work is not steadily improving the teacher's social status, the situation
becomes the primary example in the eyes of students for the fact that learning
more with more understanding is not the way to make one's life complete.
Students will learn to think independently from other sources and not from the
limited and helpless teachers.

But most importantly, a typical teacher is subjected to power of an arbitrary
educational system and she or he cannot teach students how to execute their
rights to learn, understand, build, improve, prosper and be happy in
life.\cite{Fullan} No wonder teachers believe that their work environment has
never permitted them to show what they can really do.\cite{Lortie} In turn, no
bright child wants to follow the lead to a dead end and become a teacher. This 
is very bad because without the brightest talent the teachers have little chance 
to change their status.  There is no exception to this rule, from the lowest
to highest levels of educational institutions.

Being a student and observing students one sees they do not learn at school
enthusiastically.  Students are not engaged mentally in the learning processes
to the extent required to learn effectively.  \cite{Sarason} One or two teachers
impress a student sometimes but students too rarely see their school as a source
of inspiration.  They are often forced to do things they consider useless.
Their compliance and hard work on assignments elevate their opinion among
teachers but do not bring outcomes of clear on-line value to them.  The school
does not teach us how to learn according to current needs and how to use the
knowledge and skills to steadily advance in life.\cite{Fullan}

This article describes principles one can use in attempting to change the
educational status quo at the beginning of 21st century.  The place to start is
the reader's own workplace and neighborhood.  But before the reader gets a
chance to start thinking how to go about her or his contribution to the process
of redesigning education \cite{WilsonDaviss} I need to describe the processes I
envision.  Therefore, I invent a model structure and describe how the processes
work in that structure.  This article is limited to these two subjects.  My aim
is to show one needs to start thinking in terms of the system processes in order
to find out how to make the overarching goal of education a driving force.

The vital processes of learning in the system I envision involve exchange and 
trade of knowledge, skills, materials and other resources such as time needed 
for learning and practice.  The structure of the system supports this trading.  
Nevertheless, I shall first describe the structure of the system because the 
processes of learning and trading occur in the structure.  The structure is changing
according to the needs of the processes.  Therefore, I will describe only an
initially conceivable structure.  In fact, it is not known that a single stable
structure may fulfill the needs of learning processes.  \cite{WilsonBarsky} It
is more probable that the structure will evolve as is common to life and our
civilization.\cite{deGeus}

It is essential to understand what kind of trading I am talking about.
Teachers are perceived as hired by the society to push knowledge that is not
produced by them.  Teachers are considered to be passive in creating
civilization.  Teachers are supposed to pump civilization into students' heads
as if it were pumping gas into a car.  A math teacher is perceived as someone to
teach addition, not as somebody to teach thinking and learning on the example of
addition.  A student is perceived as a car that needs gas.  Students are not
perceived as having delicate brains to be loaded with skills and checked for
flawless function by extremely competent artists of human mind-crafting.  The
way we look at the educational system resembles dumb work on trivial projects
such as repainting white boards in color or hitting white and black keys.
Students are not seen as learning how to become artists and teachers are
certainly not perceived as masters.  Therefore, teachers are denied the right to
collect money as if they were creating something desirable to us by their own
minds and hands.  Teachers are so used to the treatment that they seem not to
see they can learn more and become owners of a new generation school system.

But I need to warn the reader right away that a private school is not the idea I am
talking about.  What I am talking about is a new profession of teaching based on
serious research aiming at understanding how to teach effectively.  Professional
teachers form a network of experts on the subject matter and teaching
techniques.  The network teachers earn more if they teach better.  These
teachers can say that their profession offers them multiple opportunities for
personal and intellectual growth every day.  And the teacher future looks
brighter the more apt a student she or he is, at all levels, from the nursery to
the highest academia.

My claim is that educational systems are hard to change because teachers are not
owners of their trade.  While all prospering businesses buy something, do work
on the bought material and sell the result for a higher price, or provide
services using their own knowledge and ability which is thus being sold,
teachers are hired for a job that is considered in a way to be merely loading
wagons with potatoes.  The potatoes, trucks, ramps and trains belong to us.
Anybody can do the loading job.  And the students leave the station full of
potatoes like wagons.  We are shocked that they are not illuminated Picassos,
Liszts or Pasteurs.  We demand a lot providing merely a regular pay for
compliance with rules.

Teachers need motivation to transform their occupation into a profession.
Current systems are such that attempts to innovate put the regular income of a
teacher in jeopardy.  Trying new things is very risky because it is not
guaranteed to bring results.  Nobody wants to stop teachers from trying but the
system effectively forces them to quit because they get tired and burn out.
There is no structure to support innovation.  There is no structure to develop
innovations into elements of teaching profession and culture.  And there is no
way for teachers to be fully recognized.  Therefore, teachers have to find a way
to win the recognition of their trade themselves.  They need to start trading
with what they possess and do.

In fact, a successful teacher is an owner of incredible gifts, someone who
possesses unusual skills and knowledge and can set a high price on services
provided for students.  Students can be active learners from competent teachers.
Both need freedom to build a system to function in a natural way.  And a clear
suggestion for such a system to have a chance to succeed can be found in
universities.

University professors are considered to be owners of their wisdom because they
participate in the process of creating science.  Teachers could be seen as owners 
of their teaching materials and techniques if they were creating the materials and 
methods they use.  University students are responsible for what they spend their 
parents' money on and the same could happen with students in lower schools.  
Universities are populated by a whole hierarchy of people with different levels 
of knowledge and skills while at school we have only teachers and students.  A 
global network of schools based on the principles of a new generation university 
which belongs to teachers and students is the vision I describe.

The main problem with the model invented in this article is that it is too 
complicated to comprehend and judge quickly. Moreover, it is full of conflicts.  
If it could work it would only do so through a balance of opposing forces.  One 
can easily point out apparent inconsistencies.  To explain why the described 
system could work one would have to answer an unending chain of questions and 
the answers would stimulate new questions.  Solutions to problems showing up in 
the system have to be invented on-line as the whole system grows.  But what I am 
trying to sell is not the particular model.  I am using the model to convey the
idea that the simple principle of schools owned by teachers and students 
immediately leads to incredibly rich structures and provides clear criteria for 
distinguishing which new elements of the system might be useful and survive and 
which not and die. The ownership principle opens a new way of thinking about 
education.  If this is understood by the reader, the rest is details that may 
change in time.

\vskip.3in
{\bf 2. STRUCTURE OF THE NETWORK}
\vskip.1in

It is important not to perceive the structure I describe as rigid and ultimately
defined.  A living company evolves.\cite{deGeus} The model example I provide here 
is arbitrary but it illustrates the underlying principles. The principles 
themselves are not arbitrary. If the described model structure has drawbacks, 
including serious ones, they should be thought of as resulting from a long 
evolution and one should ask the question what processes exist in the system that 
can resolve these problems.  The structure I describe is arbitrary because no real 
process of evolution existed to create it in a natural self-correcting way.  This 
is not a problem since the model is only a tool to bring up relevant issues. 

Every teacher and every student is a member of the network as an individual with
equal rights to all other individual members.  The members form teams, classes,
schools, school districts, school regions, academies and a single society of
professors of education with its own hierarchy.  These subgroups have different
responsibilities and are distinguished by the responsibilities.

The responsibility of a team is to learn an assigned (or chosen) subject.  A
team contains about 4 or 5 members (between 3 and 7) and exists for as long as
the assignment or chosen task is not completed.  There may be 4 students in a
team, or 4 teachers, or 4 professors, or a mixture thereof.  Teams are formed
according to the demands of the subject they are supposed to study.  A team is
the basic element of the system because it is the learning engine which delivers
the result of its learning process:  a report on what and how the team learned
carrying out its project, and a product the team was supposed to produce through
the project.

Teams form classes of varying size according to the amount of assistance the
subject matter requires.  Subjects such as floating or sinking of rigid bodies
in liquids \cite{McDermott} can be studied in classes of 4, 8 or even 10 teams.
Such an ``Archimedes'' class may require one to three instructors for
assistance.  The notion of a class is distinguished by the notion of
instructors.  Since teams encounter difficulties when learning new things and
new ways of thinking they need to ask questions and verify their reasoning with 
instructors. The instructors can be students, graduate students, teachers, graduate
teachers, professors, graduate professors and professors of education.  It
depends on the kind of class they instruct.  For example, the class studying the
notion of the contexts of productive learning \cite{Sarason} according to modules
designed in analogy to Ref.  \cite{McDermott} in case of physics may require 
instructors to become members of the teams.  I will return later to the issue of 
materials for teaching the context of productive learning by inquiry since in the 
network I envision the context of productive learning by inquiry is a basic 
building element and in the current systems of education this context is almost 
entirely absent.

Classes form schools.  It is important to form schools in order to sustain
social aspects of the learning processes.  Schools are to serve their
surrounding communities.  An elementary school is fairly local in this respect
while a top ranking university may have a mission of national or even global
outreach.  Schools are distinguished by having a principal and a body of
teachers.  Schools are basic posts of the system.  Schools owned by teachers and
students form a network because it is easier to operate in the network than
outside of it.  The network is international because it draws on teaching and
learning experiences which are published and useful across the world.

Teachers form a school to create a body of sufficient expertise in the subject
matters to be able to teach and to form a setting in which their own development
of skills and knowledge will be possible through sharing duties and exchanging
experiences and results.  Teachers of one or more schools may form a team to
learn and work on some problem.  The striking feature of the system driven by
the learning of teams of various kind is a possibility of a self-organized
virtual school formed of classes of teams of teachers as a result of their own
recognition that the problem they want to study requires such structure, with a
body of super-teachers drawn from other elements of the network.

The key function the principal is responsible for is to make the school
productive.  The productivity is measured by the results of students on standard
tests and by the number of team reports from the school sold on the system
market. It is a big success to produce a good report which sells well.

We need to recall that a team may be composed not only of students but of
teachers as well, and a team may include people from outside the school.  In
that case, the team result is shared according to prescribed rules.
Consequently, the number of reports or publications or student achievements do
not need to be simple integers and instead of using special measures the school
outcome is measured in terms of money:  total sales minus total investment
divided by the number of school members.  Details of the accounting will be
discussed later, but we need to mention three things.

One is that the team results may be highly professional and even able to solve
practical community or wider problems.  Therefore, they may be copyrighted,
patented or sold.  For example, an outstanding teaching material on the subject
of sinking and floating, electric currents or optics, such as Ref.
\cite{McDermott}, may be in high demand in all schools or, a local solution to
the problem of child care and a computer program needed in its administration,
written by students and teachers, may have broad applications.

The other is that the standard test results of students need to be accounted for
in money.  Therefore, there are tables developed of equivalence between credits
and money.  Credits are universal for the whole system but they may be
equivalent to more money in a better school when considered as a product and
less money in a better school when considered as an expense.  This point will be
further discussed in the Section dealing with finances.  Here we mention only
that the accounting of schools and reviewing principal's performance in terms of
money makes it evident that education is not a burden to the society but a
source of major income if money is properly invested.  Accounting of education
in terms of money also prevents wasting public funds for education.  For it is
too easy to spend money without accountability while good accounting creates
responsibility.

The third thing is that details of the calculation do matter.  In fact, they are
essential.  It is not obvious how to evaluate results of education in terms of 
money. Therefore, the method is a subject of ongoing studies.  The studies are
essential to the network because the education it offers must be useful to the
society for the system to prosper and be actually paid as much as it aspires to.  
The studies feedback is critical to long term planning and development of
the evaluation rules for credits in terms of money.  But the studies are
essential for many more reasons, vital to the network.  Here are some examples:
design and redesign of curriculum structures, admission, examination and testing
procedures, hiring policy, communication with employers, satisfying needs of the
job market through the network and longitudinal studies of alumni careers.
Therefore, the evaluation scheme is a permanent source of initiative for
improving the network to better serve the society.  The details are hot subjects
in the network.

School districts are formed by schools spontaneously to coordinate work and
express opinions of many schools in a selective and organized fashion which
guarantees coherent action in defense or promotion of the districts educational
or other interests.  The body of representatives is elected by schools.
Therefore, the districts are distinguished by their representatives who serve
their needs.  Districts are formed to contain schools of their choice and do not
have to be restricted to primary, middle or high schools only.  Districts
prosper if their schools earn money.

School regions include school districts and universities.  The regions are
formed to allow universities and schools to utilize their resources in producing
team reports and selling them.  The principal role of regions is to provide
permanent in-service learning opportunities to every member of the system on the
highest possible level.  Teachers study in the region to keep abreast of the
science or art they teach.  University students, graduate students and faculties
study ways their research capacity can be enhanced through becoming more useful
in education, mainly through many opportunities of delegating responsibility for
teaching to students.  \cite{WilsonDaviss}

One of the students mission becomes then to work with less educated students on
their learning skills, using the best materials available.  Students who excel
in teaching can pursue studies in the system and become teachers.  It takes a
region to create conditions for such advanced studies.  Two reasons are
essential.  The region is the smallest structure whose size provides sufficient
amount of students with talent for teaching and becoming teachers of teachers.
The region is the smallest structure that can support high quality research.
Regions are large enough to create conditions for unlimited personal growth of
their members.  School regions are also useful in creating a sufficiently stable
environment to support educational processes in the periods of setbacks.

A school region is distinguished by the board of trustees whose role is to
assure healthy economy of the region educational services.  Trustees of a region
are elected by the region members.  The boards of trustees use help of
academies.

Academies are organizations quite independent of the team, school, district and
regional structure.  They are the regional networks of experts who contribute
their expertise to the region system and are recognized by the system as such.
The academies are professional organizations of providers of services to the
system.  Academies recruit their members following their own rules.  Academies
can undertake action of their choice driven by the need of the system.  A key
additional function of academies is the publication of journals.

Academies publish refereed journals on education and sell them in the system as
an additional source of income.  The new striking feature of the journals is
that they have subsections for learning materials which can be bought separately
and in large quantities. School teams may attempt to publish outstanding reports
in the journals. \cite{EJP}

The whole network of schools of all kinds requires a body of distinguished
teachers for passing judgments on issues important to the whole system.  This is
a Society of Professors of Education.  Members of the society have at least 
100,000 copies of educational materials sold through the system.  But in order 
to become a Professor of Education a candidate must have a record of working in 
the system for at least 25 years and have educated at least 25 teachers who sold 
more than 10,000 copies each of teaching materials through subsections in the refereed
academic journals.

The network of schools is global and its international character is obvious to
all members.  It is clear that translation of the academic journals plays an
important role in the international contacts and learning across the globe.

Different countries may have districts and regions of different sizes but no
administrative superstructure above regions is needed or allowed.  There exist
data banks connected in the network so that no central headquarters are required
and still the system is perfectly conscious of its identity.  Members identify
themselves by contributing to the processes of the system and by using its
structures.  Regions may easily cross state boundaries for their structure is
governed by the processes they support.  Examples of well known existing
international network structures are Internet and VISA International \cite{Hock}.

Analogies exist between the school network and other essential systems in our
civilization.  The system of electric power distribution is a leading example.
\cite{Hughes} One can think of many analogies between the two systems.  Let me
give you a surprising example.  One may think about the contexts of productive
learning as analogous to the super-conducting wires, about the processes of
learning and teaching by inquiry and circulation of teaching materials as 
analogous to the electric currents and about the overarching goal of education 
as analogous to the principle of optimizing the load factors.  Human brains are 
the sources of power.  The rules of science, democracy and total quality are 
analogous to the Kirchhoff rules.  The ownership principle is analogous to the 
closed circuit condition for the currents to flow. The transition from the 
contemporary educational systems to the networks of schools owned by teachers 
and students is analogous to the transition from the direct to alternating 
current in the case of the electric power systems.

\vskip.3in
{\bf 3. OWNERSHIP PRINCIPLES}
\vskip.1in

The forms of ownership I describe are invented for illustration, have drawbacks
and are partly contradictory.  Such a situation may be realistic but the model I
describe is not sufficiently studied to claim that much.  One would have to study
mathematical models of the ownership structure including mechanisms of
governance, income, spending and population changes to make reasonable
evaluations of the ownership principles and I have not done such studies.
Still, the ownership principles are essential to the network idea and I offer a
scenario to think about.

The system that can emerge from a real trial may evolve to other forms of the
ownership but it is clear that if the system belongs to teachers and students at
the beginning and grows successfully there will be little incentive for taking
the ownership away from the primary constituents.  And if many systems are
initiated some ownership schemes will succeed and some will die.

Two different forms of ownership exist in the envisioned network, one for
teachers and one for students.  The need to differentiate comes from the fact
that teachers support their own living (and their families') through the work in
the system while the students' living is supported by parents or other
supporters.  In addition, there is a mechanism built in for a gradual change in
the form of ownership available to students who learn particularly easily,
satisfy well defined criteria and choose to make a career in the system.

Teachers own shares.  Shares bring dividends.  Shares cannot be bought.  They
can be earned.  A teacher receives a prescribed basic number of shares when
when joins the system.  The basic number of shares ascribed to a job position is
proportional to the time necessary for doing the job and the complexity of the
tasks.  The complexity factors are tabulated and published.  The basic number of
shares corresponding to a full time job of lowest complexity brings enough
dividends to live if the whole system works productively and efficiently.
Advancing in the hierarchy of teaching positions results from the growing
ability to become responsible for more demanding jobs to which a larger
basic number of shares is ascribed.

Anyone in the system can earn more shares than the basic number for her or his
position by doing the job better.  In particular, one can earn shares by
publishing educational materials.  To give a striking example:  a cleaning staff
member of some school may publish a material on economic organization of
efficient cleaning in schools so useful it may sell in thousands of copies.  The
number of shares is proportional to the number of copies sold.  Conduction of
every activity in the network is evaluated in terms of shares.  Every teacher
knows the number of shares in the whole system and in her or his possession.
Once the yearly budget forecast is published it is straightforward to foresee
the individual basic income for the current year and everybody can evaluate
their own additional income knowing the number of shares they have in addition
to the basic number.

Students own credits.  Credits have to be earned by passing standard tests and
written and oral exams and by publishing individual and team reports through the
academic journals (this is independent of the fact that producing team reports
is the main source of learning experience for students).  To be able to earn a
credit a student has to buy a pass for a course that leads to the credit.
Students can buy passes for money.  They can also work in the system as
teachers, administrative assistants or other staff, earn shares and pay for
passes from their dividends.  When students join the system to work as
teachers their shares become sources of dividends as for teachers.

One can disclaim shares by leaving the system.  Such shares die; cease to exist.
For example, when a student finishes education in the system the number of her
or his shares at that point is multiplied by the current value of a yearly
dividend per share.  This says how much the student was making a year at
graduation - it tells potential employers how much they have to offer the alumna
or alumnus to attract attention.  In turn, students know how much they can
expect on the basis of their education.  The shares of the alumni are subtracted
from the total number of shares.  When a teacher leaves the system her or his
shares are processed in the same way.

Possession of a large number of shares opens unlimited opportunities for
personal growth and attaining intellectual freedom.  Putting the process of
individual growth of teachers and students on top of other processes and setting
priorities in such a way that ownership remains in the hands of teachers and
students no matter what happens, has one key implication:  teachers become
vitally interested in reform since reform as a process of redesign (cf.
\cite{LBR}) is the natural way of improving their own living.  At the same time
otherwise insoluble or hard problems may become less forbidding.

I give examples of problems I heard about in ``Discovery'' \cite{Discovery} and
in ``Reading Recovery'' \cite{Clay}.  ``Discovery'' and ``Reading Recovery'' are
educational reforms of unusual quality, ``Reading Recovery'' being one of the
most advanced reform models in the world.  But before I give the examples I need
to explain why I am giving these examples.  Namely, I only mean to suggest that
if the ownership by teachers and students were seriously considered from the
beginning then, new ways of approaching the problems could come to mind.
Because of my limited knowledge about ``Discovery'' (which educated several
thousand teachers) and ``Reading Recovery'' (which operates in about 9000
schools in the US alone and continues to grow), my suggestions are 
hypothetic.  However, my goal is not to tell leaders of ``Discovery'' or
``Reading Recovery'' what they should have been or should be doing.  My aim is
only to show that the ownership principle implies a new way of thinking about
problems of education.

In the case of project ``Discovery'', I think the ownership principles could
have changed the project recruiting scheme, profiles of the teacher leaders
education and organization of their work, and the motivation of teachers
attending summer institutes.  The institutes would have new elements in the
program of great interest for people seeking a status of independent thinkers
and educators.  Most importantly, however, at the end of the project when public
funds expired, the participants could be prepared to sustain their independent
network and continue to benefit thousands of students without a shock of
disappearing outside support.  The above conclusion may appear surprising in its
simplicity.  However, I recall meeting teachers, teacher leaders and directors
in the project who did not expect to be left out in the cold.  Most importantly, 
they found themselves rapidly developed, with broadened horizons and, ironically, 
unable to plainly return to their previous roles in the existing system having 
no room for growth in the directions they found attractive.

In the case of ``Reading Recovery'', one might suggest that teachers and teacher
leaders who would own the system could be motivated to develop their skills
beyond the requirements set by leading scientists.  They could have vested
interests in enhancing rather than diluting standards of their services to
children with diffusion of the system.  The research leaders could securely
develop the system if their judges were not arbitrarily selected but were mainly the
teachers and parents who know first hand the project results and appreciate
outcomes of the longitudinal studies.  Even more importantly, the research
leaders of their own self-improving system could attract new young researchers
by the fact that the principle of ``Reading Recovery'' approach to reading could
be extended to other disciplines if enough research were done.  One direction
which I consider very important is the development of materials for children 
having difficulties with learning science, materials analogous to the McDermott's 
modules on physics \cite{McDermott} and the reading books in ``Reading Recovery''.  
The projection into future would be totally unbounded and exciting.  The system 
could remain highly interesting to its leaders independently of false outside 
opinions.

\vskip.3in
{\bf 4. GOVERNANCE PRINCIPLES}
\vskip.1in

All schools in the network have equal rights and the same standards of
excellence.  For individuals, there is a schedule of ranks based on the number
of shares.  People of lower rank usually pay attention to people with higher
rank because those with higher rank usually know better how to earn shares.
The principles of effective teaching govern in the network.

The average number of shares per teacher in a school measures the quality of the
school.  Schools are also measured by achievements of their students - the
number of credits per student.

Credits are well defined through common standards and other criteria, such as
juror judgments.  The standard tests are built on the principle of one framework
problem with varying input data which imply different correct output answers using 
the same reasoning. The tests verify understanding and reasoning. The skill of 
reason is the goal of education.  The correct result is of value.

Shares become worthless if students do not buy passes to earn credits.
Therefore, teachers are interested in keeping high the number of sold passes.
This leads to the improvement of quality of education since students have
freedom to make choices how to use their budgets.  There is no prescribed
governance structure that would need to be imposed artificially because of needs
to serve other interests than the wisdom-centered learning.  \cite{KGW}

Team leaders are elected within a team.  Everybody can suggest a leader but it
is the team who decides.

Class leaders are elected by the class.  Classes can and often do bid for
teachers for specific courses.  Teachers have the right to choose with which
class they will work.  Teachers usually prefer to choose a bidding class with
highest number of credits and shares.  If there is a conflict without a rational
solution the right of choice and responsibility for making decision belong to
the teacher who has more shares in the system.

School principals are elected by the school teachers and students equally for a
period of 5 years.  There is a limit of 3 such periods for a single person to be
a principal.

District representatives are elected from the whole district membership in the 
network by all teachers and students for 7 years to ensure continuity.  Schools
vote separately on a list of candidates.  The number of votes per school is
equal to the number of shares owned by members of that school.  One person can
serve up to 2 periods.

Region trustees are elected for unlimited time.  A trustee ends her or his
service only voluntarily.  Candidates are suggested by school districts.  In
order to become a trustee one has to have at least ten times as many shares as
average number of shares per teacher in the region on the election day.  The
second condition is that the candidate for a trustee must have worked in the
system as a teacher for at least 10 years.  These conditions eliminate the
situation where some important person becomes a trustee despite that this person
is fully ignorant about the system.  This condition also helps to select people
who are successful as teachers and have a remarkable record of achievement
outside the network.

There is a danger of lowering the price of passes to sell large numbers of them.
This is easily avoided through a feedback loop because good teachers will not
work for free with too many students and such practice dies out.  On the other
hand, there is an issue of the system becoming a monopoly and dictating a too high
price on the passes.  Therefore, the system is built from more than one
independent subsystems of shares and credits.  There is no artificial limitation
on the number of such subsystems.  Teachers are free to initiate new subsystems
but there is a requirement that a subsystem must be adopted by at least three
schools.  To create a subsystem teachers of the schools set up a company
according to the common law and the subsystem becomes a partner in the network.

The standards of credit requirements and the educational materials are used
across the board equally in all subsystems.  Shares in different subsystems are
compared using the ratio of the average number of credits obtained by students
in the subsystem during the last year per share in the subsystem.  Credits are
universal because they are based on satisfying most objectively measured
requirements by students while the subsystem share value depends on the
subsystem.  Each subsystem share value is calculated in terms of a universal
share value for the whole system.  The total number of the universal shares is
equal to the sum of numbers of the universal shares in all subsystems.

All subsystems are free to function without state or community support for
students (taxes) but if they use public money as income to pay dividends they
have to comply with the general share evaluation scheme in which the share value
is defined by test results for students.  Testing schemes are continuously
redesigned to satisfy changing requirements of the job market and the network.
The testing practice is based on verifying thinking skills, understanding
subjects and ability to learn new things, in mandatory agreement with the
overarching goal of education.  That this condition is satisfied results from
the fact that the true value of the network to the society is precisely the
supply of contexts of productive learning.  In other words, the network tests
students if they purchased and acquired what they intended to when paying for
the passes.

The common share evaluation system is needed by all subsystems.  The subsystems
want to demonstrate effectiveness and quality of the education they offer.  They
want to attract the best students.  Subsystems compete by keeping the number of
credits issued per share as high as possible to keep the share value high.  The
reason standards are not reduced is that the demands for credits are universal
and in check by all constituents.  The network is also being constantly
evaluated from the point of view of the job market.  The market economy cures
negative features in the network as it does for itself outside the network.  The
bonding scheme is based on the market competition for best students and its
reluctance to hire poorly educated alumni.  In turn, nobody is interested in
educating students whom nobody wants to hire.  Therefore, the schools keep
records of their alumni careers.  There is a whole area of studies on measuring
alumni careers for meaningful comparisons.

\vskip.3in
{\bf 5. SOURCES OF INCOME AND FINANCE MANAGEMENT}
\vskip.1in

The system collects money for teaching students directly from the students budgets.
The budget money is provided, for example, by parents for education of their 
children, by a local community to pay for education of its teachers, by state and 
international organizations for training of professionals or by foundations through 
grants for scientists doing research in the system. Students buy passes to earn 
credits. 

It is essential for students to administer the process of purchasing passes.  This 
way they learn the cost of their education and how to avoid waste of their money.  
The purchase of a pass to earn every single credit is done by a student separately, 
in an on-line process of learning how to manage her or his education program.  The 
youngest students are being helped in this respect by their parents or guardians.  
There is a scheme of reducing the responsibility of parents or guardians as students 
grow up.  There is a system of consultants to students and data banks for their use.  
Every school has its own data bank with a network connection to help students make 
choices.

Students have individual budgets for their education.  All students have equal
access to the minimal budget for purchasing a basic set of passes.  The basic
set defines the level of education guaranteed to be available to every student
in the system.  Students need to raise, borrow or earn more money to cover costs
of passes to additional credits of choice for their careers.  Students
demonstrate their records to obtain such funds.

The budgets for students come from states (taxes), public or private and
national or international organizations of all kinds interested in educating
students of all kinds, and from students themselves.  But the dominant source is
the direct payment by parents or employers.  In the fully operating system which is
already blended with structures of whole societies, parents and employers may
temporarily deduct transfers made to budgets of students they support.  The
deductions are allowed for as long as a student needs to earn credits or until
the time foreseen for earning the credits expires.

However, in the current situation such a scheme is not directly implementable
since taxes are paid today according to schedules that have nothing to do with
how the tax money will be spent.  In other words, we pay taxes on what we earn
and we have no direct way to say how we want our money to be spent.  There is no
entry on the contemporary tax forms concerned with what we wish to provide our
money for, except, for example, church taxes in some countries.  Hopefully, in a
future tax forms one will find entries for education.  Until then, what I can
offer in practice today is merely a seed or initial business plan for first
steps on the way to make the educational network belong to teachers and
students.  This is described in next Sections below.

The total amount of money collected by the network for a fiscal academic year
(or semester) is divided by the number of shares issued to date and the
resulting number defines the total dividend per share.  Owners of shares decide
how to use the money they receive through their shares.  They form
organizations in the system to use their money.  For example, each and every
member of a school brings a definite amount of money for use by the school.  The
sum of money of all members has to cover all expenses of the school, including
the owners income.  A good teacher with many shares becomes thus a great asset
to the school.  Her or his opinion about teaching practice cannot be neglected.
Such teachers make the schools going.

Teachers form schools voluntarily.  If they do not form schools their shares
will lose value - no single teacher is able to offer a comprehensive education.
The same motivates formation and existence of districts and regions.

A highly sophisticated system of collecting payments, share accounts and copy
rights is in operation.  But the system rules are simple, published and easily
available.  They protect rights and intellectual properties of teachers and
students, warrant creation of the contexts of productive learning and serve the
overarching goal of education.

The financial management is not delegated to outside companies.  The outside
auditors are hired but mainly to help in eliminating errors and for
communication purposes with outside the network.

The network employs its own highest quality accountants who are also teachers.
The accountants are deeply aware of the network principles and serve well the
network educational agenda.  They teach the network accounting to their less
experienced colleagues and students.  Much of the work is done by the students
as part of their credit earning in accounting and related subjects.  Similar
delegation of work and responsibility is practiced in all administrative
functions.

The individual shares are issued by the subsystems according to their needs 
and the number of shares issued by every subsystem is decided within the subsystem.
These shares are evaluable in terms of the universal shares. The number of 
the universal shares in the whole network is an abstract number which roughly
equals the number of individual shares and results from the accounting rules. 
Thus, the individual share brings dividends in amount comparable to a universal
share. You can check what is the dividend one gets for a single share in some
school and you have an idea about the level of education the school offers.

Changes in the network accounting are induced by the majority vote, on
recommendation by representatives of districts, with approving opinion of the
region trustees.  The Society of Professors of Education is obliged to help in
assessment of proposed changes in the accounting rules used by regions.

The reason for that no monopoly can emerge is that many subsystems exist and
they compete to win their share in serving educational needs of the society.

There exist also ways of giving money to the network and specifying the money is
given to a subsystem or other unit for some purpose. The money is used then 
by issuing a corresponding number of new shares and distributing those shares 
in agreement with the intention of the donor.

\vskip.3in
{\bf 6. ASSESSMENT AND RECOGNITION}
\vskip.1in

The bottom line in assessing the quality of work and productivity of teachers
and students is the number of credits they produce.  Therefore, the credit
system is a subject of continuous research, redesign, application and feedback.
\cite{LBR} The notion of a credit guaranteed in gold explains the quality of
reasoning skills and knowledge of students who earned the credit.

The award winning teachers obtain one time money prizes or a number of shares.
Dividends from the prize shares are collectible over different periods of time;
longest times for most prestigious awards.

If the number of shares grows with time and the number of students does not grow
the value of a dividend per share becomes smaller with time, and one has to earn
more shares to keep the individual income rising.

No other measure than students performance on universal tests is used in
assessing effectiveness of teaching.  But owners of copy rights for teaching
materials and patents for teaching techniques collect royalties on their use.

Since there are differences between districts in student readiness to learn the
districts must specialize in different levels of education.  Whenever an
opportunity arises for a school district to go to a higher level the opportunity
is taken because it is preferred by the job market to employ people with higher
number of credits.  The credit system is such that gaining merely basic skills
and knowledge cannot bring a high number of credits.  A large number of credits
may be obtained only by a student who learns many skills and subjects very well
and the achieved level is verified thoroughly and trustworthy.

The protection of teachers rights to benefit from selling of their teaching
materials is secured by the general patent and copyright laws.  A recent example
of such laws are laws prohibiting unauthorized duplication of video tapes or
compact discs and fighting pirates across the world.

The highest recognition available to teachers and students is based on
leadership positions they win in their own network units.  For example, a team
distinguishes its leader, a school distinguishes its principal and academies 
distinguish their leaders.

\vskip.3in
{\bf 7. ARCHIVES}
\vskip.1in

The system keeps an archive, in many copies and in a flexible network of easy
access.  The archive is sophisticated in its purpose, structure and
availability.  Highly sophisticated librarians manage the archives and make sure
no member of the system is denied access to data.  The contemporary electronic
libraries such as the Los Alamos National Laboratory electronic preprint library
\cite{LANL} can serve as a prototype example of the network archive.

The archive plays the role of a patent office library for educational materials,
copy right guard (issuing single authorized copies), source of information and
ground for longitudinal studies (independently of the studies conducted by the
network subsystems) and the forum for research and discussion on educational
matters including the performance of the system itself.

However, the archive is not able or allowed to become a publisher or distributor
of academic journals.  The publication of journals is reserved to academies in
order to secure the journals high quality through the peer review processes.
Still, the archive is indispensable since it provides information the publishers
cannot provide, such as access to publications from different publishers.

\vskip.3in
{\bf 8. PREPARATION, DESIGN, LAUNCH, FEEDBACK AND REDESIGN}
\vskip.1in

The leading idea of the network is that teachers and students can build a
healthy and rapidly evolving educational network if they start doing it step by
step on sound economical basis.  In the envisioned system, a group of interested
teachers starts from earning funds for opening a small school.

Thus, the first step a group of teachers does is they decide they want to create
their own school.  The group then gets in touch with a local network subsystem
to learn about what and how one can do to begin with.  The local subsystem
delegates a specialist who helps the group in identifying their goals.  They
draw the first draft of their vision, mission and business plan statements to
present it to the subsystem they want to join.  The group first investment is
time and work on its own education in the already existing network.  This
education allows the group to start building a plan how to create a new school
and beyond. More about it in Section 10.

The design, launch, collecting and analyzing data and redesign processes are 
gradually becoming a habit of the group. \cite{LBR} Once they succeed in 
setting a school in operation they begin
to build partnerships with other schools of the subsystem and learn how the
network works.  The feedback from the network is essential for the new school
development.

A mature school participates in the network operation without limitations.
Schools of distinguished quality benefit from serving less developed schools by
sharing their expertise.  For example, an experienced teacher can teach a class
of colleagues how to manage their time at school more effectively and, as a
result, have more time available for personal growth.  \cite{Covey} Where is the
time coming from?  The skillful teacher knows how to help students learn on
their own, how to organize their work so that older or more experienced students
help younger or less experienced ones.  The teacher knows how to set up the
teams work so that the teacher has plenty of time to think about the most
interesting things to her or him.  Teams of students can easily do a lot of work
which otherwise overwhelms overworked and over-stressed teachers.

\vskip.3in
{\bf 9. COMMUNITY AND STATE SUPPORT}
\vskip.1in

Schools of the network are so effective they are eagerly welcome in local
communities.  It is worth for the community and state to invest in the learning
processes kept alive by the schools of the network since these processes educate
sophisticated alumni.  It is essential to understand that schools never have a
problem with getting local support because they grow out of local initiatives.
They also never lose state support because no state is going to risk opinion of
having no interest in the best possible education of citizens.  Once a school is
formed it operates for as long as its share value is reasonable.  The school
grows and brings higher income to teachers when its share value grows.

Communities are proud of the quality of their schools and press local
governments for execution of productive educational policy.

\vskip.3in
{\bf 10. SEEDS AND TIMING}
\vskip.1in

The envisioned network of schools owned by teachers and students is built in
analogy to leaving organisms.  Life begins in small seeds, not big scale
projects.  We have four elements to mention in the analogy.

\begin{itemize}
\item{A single member of the network, a teacher or a student grows from an isolated
individual of limited horizons to a member of a learning community with broad
horizons and freedom to make choices.  Thus, the seeds of the network come from
personal learning and growth of its members.}

\item{A school is born and it advances through levels of professional efficiency.  A
new post of the network emerges and supplies its strength to the whole
structure, as a leaf or root of a tree.}

\item{The whole network grows and improves its services to the society.  A small size
of an initial stage does not prevent the development into an impressive
structure as much as the size of a sequoia seed does not exclude a giant forest
in future.}

\item{The core ideas evolve from the embryos such as this article to
mature driving ideas for large systems if the ideas are helpful in practice.
The educational network is a seed for bigger changes in our global civilization.}
\end{itemize}

The small size of seeds is important, not accidental or resulting solely from
financial limitations.  The small size of seeds is the condition which
eliminates huge errors.  The hardships of life teach members of the network how
to go about developing their own schools.  They are helped by the network in a
number of essential ways characteristic of productive teaching but a school must
learn how to grow and become strong.  Only self-consistent and well adapting
structures survive.  Thus, even if the network could afford financing large
scale projects it does not do it lightly without careful studies.

It will take a long time before a reasonably mature version of the envisioned
network can become a reality and show its ability to improve and survive on the
basis of effective education of its members. 50 to 100 years is not a bad guess.
Before that happens, contemporary groups of teachers interested in building such
a network must start with other enterprises in mind than creating the whole new
educational system at once.  The seeds must be sown differently.

The point is following.  The existing educational systems have long traditions.
There exists no counterexample of comparable magnitude that could substantiate
claims one can do better.  It is pointless to quarrel about what is good or bad
and engage in easy to create wars of opinions.  Moreover, it would be silly and
immature on anybody's part to claim knowledge of how to build a better system of
education than the existing ones to the extent that once the solution is adopted
from top to bottom no problem will arise.  The principles of life do not favor
concepts such as Frankenstein.  Therefore, the network concept stays away from
such ideas.

In contrast, teachers and students undertaking action to address their
basic needs to learn and grow should not be met with a strong opposition.  For
opposing such movement cannot be supported by reason and any such opposition
would contradict the purpose of education.

The timing idea is that once a growing self-organizing network of teachers and
students begins to practice meaningful education it will be easier to found,
fund and find (3f) exemplary schools where the contexts of productive learning
\cite{Sarason} and learning by inquiry \cite{McDermott} is bread and butter.

Teachers need to prepare themselves to take initiative much earlier.  One place
to start with is after-school or after-work activities for youth and adults in
the local community.  \cite{Ekiel-Jezewska} The author knows that physical
phenomena such as electric currents flowing in a circuit of batteries, bulbs,
switches and wires, or a daily movement of the gnomon shadow on a sundial,
provide opportunities for teaching how powerful is learning by inquiry.
Understanding of the solar system or laws of electricity on the basis of a
conscious inquiry induces deep changes in the learning habits.  To get going, a
team of teachers needs to see this happen to themselves and their students.
Then, they need to repeat the success working with new people.  For example, one
can teach grandparents how to work with their grandchildren.  Another viable
program is a summer vacation or holiday camp for youth or adults.
\cite{Ekiel-Jezewska} It is essential that the activities are conducted by
experienced people using materials of high quality, at least as good as
``Physics by Inquiry'' \cite{McDermott}. Teachers engaging in such initiatives
become active learners of the subject matters.  More importantly, they begin to
learn how they can become professionals.  Teaching and learning according to
principles of scientific inquiry are the cornerstone processes in the network
development and one has to begin there.  The same principles are then available
for application in teaching and learning in other areas without limitation.

The initial studies must be economically sound, with direct collection of money
from parents or employers.  This way first self-organized learning teams of
teachers may emerge. Otherwise they burn out. The teams learn what is involved 
in the enterprise.  For example, a small company is subject to many laws teachers 
do not learn about in college.  \cite{Markowski} The small scale operation is 
an indispensable source of knowledge for teachers about what they can accomplish.  
Only those who know their trade can build a school of the network.

The key characteristic of the contemporary situation in educational systems is
the lack of shared understanding what is the goal of education.  The seed
activity must focus on building a shared practical vision of education among a
team members.  The remaining paragraphs in this Section describe the first seed
activity which the founders of the network should seriously consider.

My claim is that the notions of the context of productive learning, learning by 
inquiry and the overarching goal of education are not commonly understood.  I 
have noticed in science and educational institutions that almost never the 
bottom line of research and learning is put on the table as worth investigation.
I have not heard people asking seriously what and how we really want to learn 
and why.  Such issues go beyond the common discussion. They seem to be obvious.  
I claim they are too damn difficult to understand so that anybody worrying about 
their position cannot seriously admit ignorance in this area.  The ignorance is
covered by a tendency to push in the direction which is known and safe to the
individual.

The next logical step is to ask:  Do we have a textbook, a module analogous to
Ref.  \cite{McDermott}, and a program which would be teaching what is the
essence of productive learning by inquiry?  My answer is:  No.  Moreover, the
non-existence of such a module for learning how to teach by inquiry clearly
shows our reform efforts are weak and missing the key innovation element.  I
suggest that the teams interested in building the network of schools owned by
teachers and students start from creating their own versions of such modules.
The modules should then be refined in the process of educating new people who 
join the founding teams.  A new school founding team should start from work on 
their school cooking-book and I claim they should start from learning themselves 
what is food.

I give you another analogy which is useful here.  Think about the educational
system and the air transportation system.\cite{WilsonDaviss} In the air
transportation the notion of flying was clear to everybody since they saw a
smallest bird in action.  From Icarus to Boeing 747, all participants of
whatever was being done knew beyond any doubt what they have to demonstrate or
see in order to say they fly.  In the educational system, no analogous notion
exists.  The notion of flight in education can be the context of productive
learning by inquiry, but it seems to be top secret now, or more probably, the 
notion does not even exist in the most educators and politicians minds.  I do not 
blame anybody. The notion is very difficult to understand. You have to combine 
ideas of most advanced sciences and arts, climb high up on top of them and see 
far enough to come to grips with the notion of flying in education to try your 
own wings. Talking about the design of the wing curvature or the airplane 
factory management is premature when the notion of flying is not known.  A school 
founding team needs to understand what they mean by flying in education 
before they can start working on their propeller.

Moreover, the majority of complaints about performance of the educational
systems is not serious.\cite{Fullan} Namely, the relevant people do not
understand the crisis to the extent of saying:  I am not doing the right thing
now, I am not teaching effectively, I do not know what is the context of
productive learning by inquiry, I am unable to achieve the overarching goal of 
education for my students, I, in the first place, need to start thinking what 
I am doing. In other words, it is not only unclear what is flying but it is 
also not true that people realize they do not know what is flying in education.  
To the contrary, most educators are convinced they know something well enough
to teach others. In fact, it often becomes comparable to teaching Little Mermaid 
how to comb her hair with a fork.\cite{Disney} The way one shows to somebody 
that something is not the way that person thinks is following:  one asks the
person to make a verifiable prediction of the real status of the matter in 
question and then one verifies the prediction together with the person.  If 
the verification shows the person's prediction was false the person is shocked, 
becomes curious, starts thinking begins to listen. This is teaching by inquiry.  
To start learning by inquiry you have to feel safe to ask questions about 
what bothers you.

My punch line is here.  There is a science of incomparable clarity and focus in
learning by inquiry.  It is physics.  Basic physics is the most transparent source
of understanding what it means to learn effectively.  And we already have a well 
tested material to study the notion of learning by inquiry in physics.  This is 
Ref. \cite{McDermott}.  The first thing to do for a team of teachers who want to
understand what they are truly after if they want to join the living network of 
schools is to study electric circuits or optics in the way similar to the one 
from Ref.  \cite{McDermott}.  Then comes learning about what ``Discovery'' 
accomplished in Ohio and where it failed.\cite{Discovery} At this point one 
begins to understand the value of the context of productive learning in physics
and how hard it is to create it. Then, one needs to ask what is ``Reading 
Recovery'' extending from New Zealand to the U.S.A. and where it is 
going.\cite{Clay} The context of productive learning by inquiry comes at this
stage more clearly into your sight. Next step is to talk to the Learning by Redesign
\cite{LBR} and learn by inquiry in the context of your path about past reforms 
and the status of science of change. Finally, once you become an owner of a clear 
notion of the context of productive learning by inquiry you can start your
independent thinking about the living network of schools owned by teachers 
and students. If this outline frightens you, forget the network idea.

%\newpage
\vskip.3in
{\bf 11. INITIAL BUSINESS PROPOSAL}
\vskip.1in

The faculty member in a university who is prepared to do so might help a few
colleagues to study the notion of productive learning by inquiry and understand the
benefits of using the technique.  A group of faculty members could set a program
for learning by inquiry in the areas they see fit, at all required levels and
for all people they want to talk to.  This activity could produce the first
shared notion of the purpose of education in the faculty team.

Three school teachers and a university faculty could spend a semester preparing
a one semester course on electricity or optics by inquiry.  They would have to
provide their money for equipment and their time for the work and learning.  In
Poland, the investment could presumably be at the level of about 150 z{\l} per
person per month.  That makes 2400 z{\l} for 4 people in a 4 month semester with
about 2 hours of studying session and 2 hours of preparation a week.  The team
would learn principles of learning and teaching by inquiry.

In a second semester, the same team could deliver a course for youth, adults, or
both, to about 30 clients for money.  Each client paying 50 z{\l} a month makes
1,500 z{\l} per month and about 6,000 z{\l} in four months.  Divided by 4 it
gives 1,500 z{\l} per teacher for all expenses of the course, or 375 z{\l} per
month.  Suppose all the money is spent on equipment and other expenses.  One can
buy a lot of nice stuff for this money.  The next semester has much lower
spending because a lot of the equipment is already in place, the time required
is shorter since there is experience accumulated, and the price may even go up
if the first trial creates demand and the course is considerably improved.
Suppose the spending is slightly higher than in the first semester of
preparation, say 175 z{\l} instead of 150 z{\l}.  Then, the next semester brings
200 z{\l} per month of income per teacher.  This means the income compensating the
initial money investment in one or two years.

The major product of the first few semesters of the team work is a set
of materials which allow a skillful teacher to engage many students in
a highly productive learning process. Such material can be published and
sold in the future in many copies. But what is created goes far beyond that
- teachers begin to act using their new skills and their own process of 
continuous learning and self-development takes off.

The question is how to move forward with the idea of founding a new school.  One
team is not sufficient.  Organization of summer camps and new courses should
lead to a larger group of teachers in association with university teachers who
can conceive a mission, a vision and a plan to found the school.  A breathtaking
variety of problems need solution to make this work.  Business strategies are
just a small part of it (for example, see Ref.  \cite{Markowski}).  But it is
hard to imagine anybody or anything will be able to stop the development.  On
the contrary, with such grass root movement and solid preparation one can expect
many foundations of education to be ready to support the plan.  Publishers and
distributors of the teaching materials would certainly be interested in the
promotion and the widest possible use of maximal number of issues.  They would
sign contracts for producing such materials.

This time my punch line is here:  the university faculty trained in teaching by
inquiry could start teaching elementary science by inquiry to teachers whose
participation in the program would be paid by employers (schools,
communities, institutions engaging in education of teachers, foundations).  The
great benefit to the university faculties is they suddenly become obviously and
undeniably useful to the whole society (remember, students have parents and it is
the parents who keep the country going), irreplaceable and in high demand for doing 
what they are well prepared to do subject-wise, what they enjoy to do by nature
and what they can eagerly do to raise their income without interference
with their research as much and as stressfully as teaching ineffectively hoards
of under-prepared students interferes with their underpaid research activities at
the university.  Long term benefits to scientists are then obvious and not
limited. Most importantly, however, the faculty members begin to feel free to 
learn a completely new stuff and start thinking in new dimensions.

There is an important aspect to mention in such an approach:  those who learn by
inquiry are ready to tackle hard problems.  They will spontaneously advance
their knowledge and understanding.  They will be driven by curiosity.  They will
create the culture able to sustain the movement towards the network of schools.
And they will grow personally with the development of the network.  The network 
will serve more students and bring respect and enthusiasm to leaders of effective 
learning. The network will continuously need its top experts to keep it on track
and going. Everybody will have to learn at new levels about new difficult matters 
and how to solve problems efficiently, bringing splendor to the educational 
enterprise.

New teams will be trained to become able to offer training to many new teachers.
The university faculty sharing the vision will be able to help in building the
network of schools owned by teachers and students and the schools will produce
students ready to study at the universities.  Training of new teams will be based on
a set of meta-teaching materials on the subject matter and methods of founding
new schools and developing the network.  People will not be merely hired to do
all these things - they will own the network and make it live up to their
expectations.  The network will become a good client of the university faculties.

\vskip.3in
{\bf 12.  COMMUNICATION}
\vskip.1in

The key roles of communication are exchange of information within the network
and informing society about the network ability to teach, improve itself and
grow.  Internet-like structures may help but they will be only tools in the
processes of importance.

The most important communication process in the network is the transfer
(describing, explaining, selling and buying) of teaching materials combined with
courses on teaching and learning how to use them offered by specially trained
users.

The most important information sent to the society is the current dividend per
share and the total number of existing universal shares.  The published numbers
include also the average number of shares earned by teachers and students in the
whole system.  In addition, tables of subsystem averages are published with
explanation of their meaning.  The tables help students, parents, foundations
etc. in judgment of the subsystems performance.  Next, the cost of the education
of a single student and the profit from education of a single student are
published with explanation of how both are calculated.

\vskip.3in
{\bf 13. CONCLUSION}
\vskip.1in

There is one consequence of the educational system belonging to teachers and
students at all levels that has not been fully described yet and must be
reiterated here.  Namely, such a network could naturally support basic research.
Moreover, it could do so without asking for immediate industrial applications.
The new motivation comes from the fact that an educational system will not be
truly useful, indispensable and always worth investment unless it becomes an
independent source of enlightenment.  New discoveries could first apply in
driving education before being used in industry.  Today, we learn at schools 
what happens in the world.  In the new system, it would be natural for the 
world to eagerly learn what is being discussed at schools - an unthinkable
situation today.

While today no economic competition in scientific progress between educational
systems and industries is possible, the new system could engage in such
competition.  The engagement might have incredible consequences for the speed of
developing our civilization.  Imagine young people learning about the current
status of our knowledge and understanding of the real world first hand and
searching for solutions to problems without bias of employment and other
commitments.

I also need to explain the opportunity the network creates for the contemporary
university.  The unique opportunity lies in the leadership role the university
could try to attain.  But we need to remember that the contemporary university
is not an unquestionable institution that fulfills its mission and may securely
keep going as it did so far.  \cite{Hock}

The well known problem the university faces is that freshman students are not
sufficiently educated at schools to undertake studies of modern science.  The
university becomes a place to teach elementary subjects because schools cannot
fulfill their mission.  Schools are supposed to teach so much so quickly that
they are unable to help students learn with understanding.  Understanding is
replaced with a mindless drill of memory.  Students become alumni who do not
know how to learn new subjects.  Worse, they are trained in faking knowledge and
understanding.  Thus, the contemporary university must face the highly probable
possibility of becoming a high school of the 21st century and never gain the 
leadership position it might dream about or believe in attaining.

The main point is not merely that the university teachers would certainly enjoy
having better prepared students as entrants (better students means better
chances for prosperity of their professors).  The point is that the university
may become obsolete and useless no matter how good a science it supports if the
students will not be able to study there in sufficient numbers.  The above
statement is not guaranteed to motivate a revolution in educational 
paradigms.\cite{Kuhn} But it means that universities have to help schools in 
changing their practice.  I give the following example.

The university works as a hierarchy of teaching and learning staff from students
to graduate students to teaching assistants to postdocs to levels of
professorship, and administration.  Climbing the ladder is related with
achievements in science, teaching, building research teams, personal growth,
gaining respect and winning intellectual freedom without limits.  Simpler tasks
are delegated down the hierarchy.  The most advanced processes of study and
teaching at the university are in the hands of the most talented and most
educated people.  How is the school organized?  There are only teachers and
students, and administration.

The questions to ask at the university are the following.  How would you explain
the utility of your system to school teachers if they asked?  How could one
implement similar principles at the school level?  Why do teachers not come to
ask how to do that with their students?  At the same time one could ask the 
following questions at school.  Why don't you try to create a structure like 
in a university?  Why don't you talk about it with the university people?  
Are they not helpful or plainly ignorant?  The university should consider the 
opportunity of helping to build and lead a living network of schools owned by 
teachers and students.  It could make a difference.

%\newpage
\vskip.3in
{\bf Acknowledgment}
\vskip.1in

The author would like to thank Ken Wilson for many stimulating
discussions and comments.  Multiple discussions with Seymour Sarason are
gratefully acknowledged.  The author benefited from meetings with
Charlie Ericson, Constance Barsky and Ben Daviss.  He also wishes to
thank Maria Ekiel-Je{\.z}ewska for discussions and collaboration on
teaching experiments.

\end{document}